\documentclass[12pt]{article}
\usepackage[utf8]{inputenc}
\usepackage[english]{babel}
\usepackage{amsmath, amssymb, graphicx}
\renewcommand{\theequation}{\arabic{section}.\arabic{equation}}
\usepackage{hyperref}
\usepackage{braket}
\usepackage{color}

\newcommand{\be}{\begin{equation}}
\newcommand{\ee}{\end{equation}}
\newcommand{\bea}{\begin{eqnarray}}
\newcommand{\eea}{\end{eqnarray}}

\newcommand{\fnm}{\footnotemark}
\newcommand{\fnt}{\footnotetext}

\begin{document}

\begin{center}

  \large \bf
Wilson loops in exact holographic RG flows at zero and finite temperature.
\end{center}

\vspace{15pt}

\begin{center}

 \normalsize\bf
         Anastasia A. Golubtsova\fnm[1]\fnt[1]{golubtsova@theor.jinr.ru}$^{, a,b}$
       and Vu H. Nguyen\fnm[2]\fnt[2]{nguyenvu@theor.jinr.ru}$^{, a,c}$

 \vspace{7pt}

 \it (a) \ \ \ \ Bogoliubov Laboratory of Theoretical Physics, JINR,\\
Joliot-Curie str. 6,  Dubna, 141980  Russia  \\

(b) \ Dubna State University, Universitetskaya str. 19, \\
   Dubna, 141980, Russia

(c) \ Institute of Physics, VAST, 
10000 Hanoi, Vietnam

 \end{center}
 \vspace{15pt}

\begin{abstract}
We study rectangular timelike Wilson loops at long distances in the exact renormalization group flow in the context of the holographic duality. We consider the 5d holographic model with an exponential dilaton potential constructed in [\href{http://arxiv.org/abs/1803.06764}{{\tt arXiv:1803.06764}}]. To probe the RG flow backgrounds at zero and finite temperatures  we calculate the minimal surfaces of the corresponding string worldsheets. We show that a holographic RG flow at $T=0$,that mimics QCD behaviour of the running coupling, has a confining phase at long distances, while the RG flow with an AdS UV fixed point is non-confining.

\end{abstract}

\section{Introduction}
In the recent years the holographic correspondence \cite{JMM}-\cite{Witten} has been shown to be a very useful tool to get theoretical insight into systems  in the strongly coupling regime \cite{CSLMRW,IYa1}. In the holographic prescription renormalization group flows of field theories are realized as gravitational domain-wall solutions interpolating between boundaries \cite{deBoer:1999tgo}-\cite{SLN}. 
These boundaries of the solutions are supposed to be related to fixed points of a dual field theory and should have particular properties.
The radial position in the bulk geometry corresponds to the energy scale, 
while the dilaton is related  to the running coupling\cite{Emil}. Various non-perturbative phenomena that are inaccessible in the dual field theory can be captured by the holographic RG flow \cite{GK,GKN}.
There is  a class of useful bottom-up holographic models \cite{HolVen1}-\cite{HolVen2}, that are of interest because they can be studied analytically  resembling features of the real QCD, in particular the running coupling \cite{AR,AGP}.

In \cite{AGP}  exact holographic RG flows for a model with the dilaton potential which represents  a sum of two exponential functions were constructed. From one side the choice of the potential is motivated by applications to studies of late-time behaviour of the quark-gluon plasma \cite{GJP}.  Particularly, a single exponential potential, used in  \cite{GJP}, corresponds to the IR limit of the improved holographic QCD potentials \cite{GK,GKN}. The case of the dilaton potential with two exponential terms was also considered in \cite{Umut}.

 Secondly, the motivation for the model comes from  a possible supergravity embedding where exponential terms can be related with the contribution of $0$-th forms \cite{9511203}. In the context of the holographic RG flow the gravity model with simple exponential was discussed in \cite{KLN}. The model  in \cite{AGP} appears to be solvable and can be reduced to  $A_{1}\times A_{1}$ Toda chain. It was shown that  the behaviour of the running coupling $\lambda$ built on the analytic solutions can mimic that one we see in QCD. Namely, one can observe an infinite value of $\lambda$ in the IR region and an asymptotic freedom in the UV limit. Moreover, for the finite temperature case the dependence of the free energy on the temperature signalized the Hawking-Page phase transition. Hereby it is of interest to perform a check if the confinement takes place for the  backgrounds found in  \cite{AGP}. 

The Wilson loop represents one of criteria for studies of the IR region and manifests an area law behaviour in the case of the confinement. Via the holographic approach Wilson loops has been widely used to consider in different settings. Following \cite{Malda}-\cite{BISY} the holographic Wilson loops can be obtained minimising the world-sheet action of a string living in the gravity background with endpoints on the boundary contour. Wilson loops in various holographic RG flows were studied in \cite{DisZam}-\cite{CSH}. In \cite{DisZam}-\cite{PZ} Wilson loops were discussed in holographic renormalization group flows in the framework of 5d gauged supergravities. 
In \cite{AR,ARS} holographic Wilson loops with different orientations were studied in the holographic RG flows constructed for the spatially anisotropic model with quadratic dilaton and Maxwell fields \cite{AR}. It was shown in \cite{ARS} that the holographic model has a confinement phase and the phase diagram  includes both the Wilson transition and Hawking-Page phase transition lines, which can have a common point. It is interesting to note that the dilaton potential from \cite{AR}, which is restored for their solutions, has the same shape as the dilaton potential considered in \cite{AGP}. In \cite{CSH} it was proposed a general method of Wilsonian renormalization to a string stretched in the holographic RG flow background.

In this work we shall consider holographic time-like rectangular Wilson loops for the backgrounds found in \cite{AGP} that can be associated with the holographic RG flows. We consider both the case of zero and finite temperature. 
Using the method  developed in \cite{AGP,Arefeva}, based on introducing a so-called effective potential $V_{eff}$, we estimate the behaviour of holographic Wilson loops at long distances. It is supposed that the confinement phase takes place when $V_{eff}$ has a critical point. We show that $V_{eff}$ corresponding  to the holographic RG flow at $T =0$, which running coupling mimics the QCD behaviour, can indeed have a critical point under certain constraints, i.e. this background has a confinement phase in the IR region.

The structure of the paper is organized as follows. In Sec.~\ref{Sec:Overview} we give an overview of the model and the corresponding generic solutions. We also briefly discuss the asymptotic behavior of the solutions at $T =0$. In Sec.~\ref{Sec:WLT0} we explore holographic Wilson loops at zero temperature. In Sec.~\ref{Sec:WLTf} we describe the black hole solution in short and estimate the behaviour of the Wilson loops in the holographic RG flows at finite temperature. In Appendix~\ref{appendix} we present the useful relations for extremal points of $V_{eff}$ in the confining background at $T=0$.

\section{The setup}\label{Sec:Overview}
\subsection{The holographic model}
Our starting point is a 5d holographic model \cite{AGP} governed by an action of  the form
\begin{eqnarray}\label{1.1}
\mathcal{S} = \frac{1}{16\pi G_{5}}\int d^{5}x \sqrt{|g|}\left(R  - \frac{4}{3} \left(\partial \phi\right)^{2}  - C_{1}e^{2k_{1}\phi} - C_{2}e^{2k_{2}\phi}\right) + G.H.,
\end{eqnarray}
where $C_{i}$, $k_{i}$ with $i = 1,2$ are constants. We choose the constants $C_{1}$ and $C_{2}$ as follows
\bea\label{1.1b2}
C_{1}<0, \quad C_{2} >0.
\eea
Putting the constraint on the dilatonic constants as $k_{1} = k$,  $k_{2} = \frac{16}{9k}$,
with $0<k<4/3$, one can reduce the model (\ref{1.1}) to the $A_{1}\times A_{1}$ Toda chain \cite{AGP}. 
The  general form of  the solutions for EOM following from (\ref{1.1}) is given by the metric
\bea\label{MABCnv}
ds^{2}=  F^{\frac{8}{9k^{2} -16}}_{1}F^{\frac{9k^{2}}{2(16-9k^{2})}}_{2} \left( - e^{2\alpha^{1}u}dt^{2} + e^{-\frac{2\alpha^{1}}{3}u}d\vec{x}^{~2} \right) +
F^{\frac{32}{9k^{2}-16}}_{1}F^{\frac{18k^{2}}{16-9k^{2}}}_{2}du^{2},
\eea
where $\vec{x} = (x_{1},x_{2},x_{3})$ and the corresponding dilaton 
\bea\label{FD.4nv}
\phi =  -\frac{9k}{9k^{2}-16}\log{F_{1}} +\frac{9k}{9k^{2}-16}\log{F_{2}}.
\eea

The functions  $F_{1}$ and $F_{2}$ in (\ref{MABCnv})-(\ref{FD.4nv}) are given by
\bea\label{nC1pC2.F1E1n}
  F_{1} &= &\sqrt{\left|\frac{C_{1}}{2E_{1}}\right|}\sinh(\mu_1\,|u-u_{01}|),\, \mu_1= \sqrt{\left|\frac{3E_{1}}{2}\left(k^{2}-\frac{16}{9}\right)\right|},\\
\label{nC1pC2.F2E2p}
F_{2}& =& \sqrt{\left|\frac{C_{2}}{2E_{2}}\right|}\sinh(\mu_2\,|u-u_{02}|),\,  \mu_2 = \sqrt{\left|\frac{3E_{2}}{2}\left(\left(\frac{16}{9}\right)^{2}\frac{1}{k^{2}} - \frac{16}{9}\right)\right|},
\eea
where  $0< k <4/3$, $C_{1}$ and $C_{2}$ are given by (\ref{1.1b2}).

We note that the constants of integration are related by the zero energy  constraint
\bea\label{E1E2c3}
E_{1} + E_{2} + \frac{2(\alpha^{1})^{2}}{3} = 0, 
\eea
with $E_{1} < 0$, $E_{2}>0$.

So the general case gives  two classes of the solutions, namely, the solutions that possess the  Poincar\'e invariance with the parameter  $\alpha^{1} =0$
and those solutions with $\alpha^{1}\neq 0$, i .e. that doesn't have the Poincar\'e symmetry.
 From the solutions with the broken  Poincar\'e symmetry we can construct regular black holes, that the parameter $\alpha^{1}$ can be related to the Hawking temperature.

Moreover, we also have constants of integration $u_{01}$, $u_{02}$, that split our solutions into three charts : {\bf a)}$u<u_{02}$; {\bf b)}$u_{02}<u<u_{01}$; {\bf c)}$u>u_{01}$;
we also aim to consider the  degenerate case with  $u_{01}= u_{02} = u_{0}$. This special case leads to the vacua with the AdS boundary in the UV limit.

We note that all branches of solutions with $\alpha^{1} = 0$ are the Poincar\'e invariant  domain walls, 
but  not all of them can be interpreted as holographic RG flows.  This can be seen from the behaviour of a scale factor of a certain background. Here it's convenient to come to the domain wall coordinates, thus the generic metric takes the form
\bea\label{domainwall}
ds^{2} = e^{2\mathcal{A}(w)}\left(-dt^{2} + d\vec{x}^{2}\right)  + dw^{2}.
\eea
We note that the change of the coordinates is given by
\bea\label{dwtr}
dw &= &F^{\frac{16}{9k^{2}-16}}_{1}F^{\frac{9k^{2}}{16-9k^{2}}}_{2}du,
\eea
so the scale factor is
\bea\label{scaleE-dw}
{\cal A}&=&\frac{4}{9k^2-16}\,\log F_1+ \frac{9k^2}{4(16-9k^2)}\,\log F_2.
\eea
The quantity $e^{\mathcal{A}}$ is supposed to be associated with the energy scale of the dual theory, while $e^{\phi}$ is related to running coupling. So it supposed that $e^{\mathcal{A}}$ should be monotonic.

\begin{figure}[h!]
\centering
 \includegraphics[width=4.6cm]{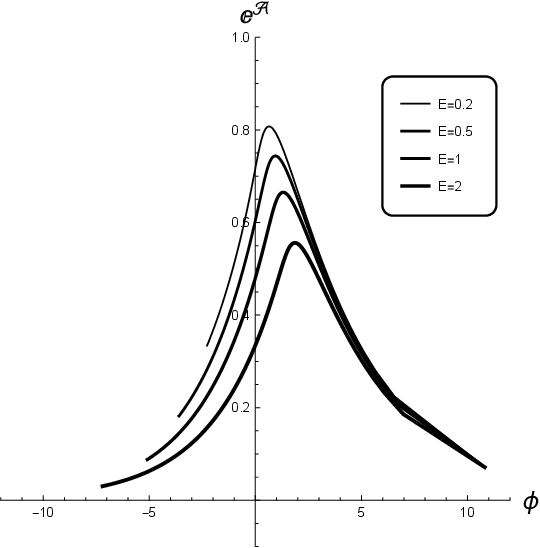}{\bf a}$\,\,$
 \includegraphics[width=4.6 cm]{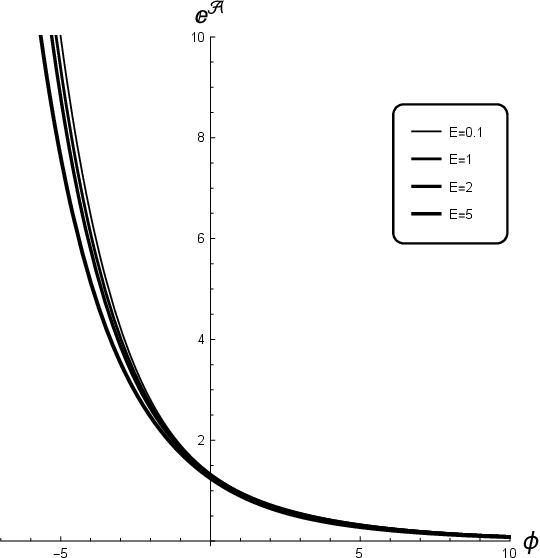}{\bf b}$\,\,$
  \includegraphics[width=4.6 cm]{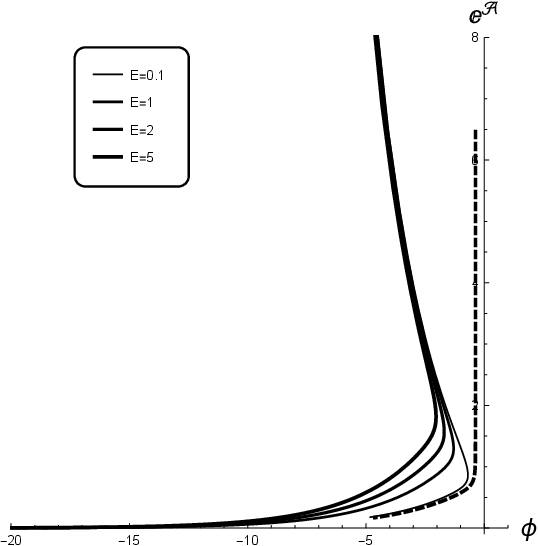}{\bf c}$\,\,$\\
 \caption{The behaviour of $e^{\mathcal{A}}$ as a function on $\phi$ for solutions defined: {\bf a)} on the region $(-\infty,u_{02})$ with $u_{02}=-1$;  {\bf b)} $(u_{02},u_{01})$ with $u_{01}=0, u_{02}=-1$;  {\bf c)}   $(u_{01},+\infty)$ with $u_{01}=0$  by solid curves, the solution with $u_{01}=u_{02}=0$ by the dashed curve for  $E_{1} = -E_{2}= 0.01$. For all plots we fix the potential $k=1$, $C_{1}= -C_{2}=1$, varying values of $E_{1} = -E_{2}$ (labeled as $E$ on the plot), except the solution {\bf c)} defined on $(u_{0};+\infty)$.}
  \label{fig:scalefactor}
\end{figure}

From figure~\ref{fig:scalefactor} {\bf a)} we see that $e^{\mathcal{A}}$ plotted for the left solution is non-monotonic, and therefore it cannot be interpreted as a holographic RG flow. In what follows we skip the discussion of this solution and its application. At the same time, one can observe from figs.~\ref{fig:scalefactor} {\bf b)} and {\bf c)} that $e^{\mathcal{A}}$ behaves monotonically both for the middle and the right solutions. The case with an AdS boundary  has also the monotonic scale factor as it can be seen from Fig.~\ref{fig:scalefactor} {\bf c)}.

\subsection{Boundaries of the holographic RG flows at zero temperature}
Here we give a review of the asymptotics of the middle and right solutions with $\alpha^{1} = 0$. As we said before, these are Poincar\'e invariant domain wall solutions,
 that can be considered as holographic renormalization group flows at zero temperature.

{\bf The middle solution}  is defined on the region $(u_{02},u_{01})$. 
The metric and the dilaton near the boundary with $u \to  u_{02} +\epsilon$ are
\bea
ds^{2} &\sim& (\mu_{2}(u - u_{02}))^{\frac{9k^{2}}{2(16-9k^{2})}}(-dt^{2} + d\vec{x}^{2}) + (\mu_{2}(u - u_{02}))^{\frac{18k^{2}}{16-9k^{2}}}du^{2},\\
\lim_{u\to u_{02}-\epsilon} \phi &\sim&-\frac{9k}{16-9k^{2}}\log\left(\sqrt{\frac{C_2}{C_1}}\frac{\mu_2\,\epsilon}{\sinh(\mu_1\,(u_{01}-u_{02}))}\right) \to +\infty.
\eea

Taking into account the plot ~\ref{fig:scalefactor}~{\bf b)} we see  that with the dilaton going to $+\infty$  
the energy scale $e^{\mathcal{A}}$  tends to $0$, so one can associate this boundary with the IR region.
We note that this boundary is a conformally flat spacetime,  the sign of the scalar curvature depends on the choice of the shape of the potential, i.e. the value of the dilaton coupling $k$ \cite{AGP}. \\
Another boundary of the middle solution is this one near $u_{01}$. Here the metric reads
\bea\label{u01ds2}
ds^{2} &\sim &  (\mu_{1}(u - u_{01}))^{\frac{8}{9k^{2}-16}}(-dt^{2} + d\vec{x}^{2}) + (\mu_{1}(u - u_{01}))^{\frac{32}{9k^{2}-16}}du^{2}, 
\eea
and the dilaton is 
\bea\label{u01dil}
\phi_{u\to u_{01}+\epsilon} &\sim&-\frac{9k}{16-9k^{2}}\log\left(\sqrt{\left|\frac{C_2}{C_1}\right|}\frac{\sinh(\mu_2\,(u_{01}-u_{02}))}{\mu_1\,\epsilon}\right)
\to -\infty.
\eea
This boundary is also conformally flat spacetime spacetime with the sign of the scalar curvature depending on the value $k$.
 Looking on the plot \ref{fig:scalefactor}~{\bf b)} we see that  the energy scale corresponding to the middle solution goes to $+ \infty$ as the dilaton tends to $-\infty$. Thus, one can say that the boundary near  $u_{01}$ is the UV limit and we see the asymptotic freedom near it.\\

Now we turn to {\bf the right solution} that is defined for the region  $(u_{01}, +\infty)$. The first boundary of the right solution is near $u_{01}$ point and the asymptotics of the metric and the dilaton coincide with those for the middle solution  (\ref{u01ds2})-(\ref{u01dil}). With respect to Fig.~\ref{fig:scalefactor} {\bf c)} we can also say that this boundary corresponds to the UV region.\\
For the other boundary with  $u \to  +\infty$ the metric reads
\bea
ds^{2} &\sim& e^{-\frac{2\mu_{1}u}{4+3k}}\left(-dt^{2}  + d\vec{x}^{2} \right) + e^{-\frac{ 8\mu_{1} u}{4+ 3k}}du^{2},
\eea
and the dilaton
\bea
\phi_{u\to+ \infty} & \sim& -\frac{9k}{16-9k^2}\Big[(\mu_2-\mu _1)\,u+\frac12\log  |\frac{C_2}{C_1} |\Big]\to -\infty.
\eea

From Fig.~\ref{fig:scalefactor}{\bf c)} we see that the scale factor goes to $ -\infty$ as the dilaton tends to $-\infty$ and we observe the IR free theory. We remind that the dilaton tends to $-\infty$ near the other boundary is well so the dilaton is bouncing.
We note that the boundary of the right solution with $u\to +\infty$ is also a conformally flat spacetime with the positive curvature.\\

{\bf The special case of the solution} defined on the region $(u_{0},+\infty)$ with  the  points $u_{01}= u_{02} = u_{0}$   is of particular interest.

The metric near the boundary with $u \to u_{0} + \epsilon$ tends to be a 5d AdS spacetime
\begin{equation}\label{5dAdSu}
\begin{split}
ds^2&\sim \left(\frac{4}{3k}\right)^{\frac{9k^2-16}{2(16-9k^2)}}[\mu_1(u-u_0)]^{-\frac{1}{2}}(-dt^2+d\vec{x}^2)+\left(\frac{4}{3k}\right)^{\frac{18k^2}{16-9k^2}}[\mu_1(u-u_0)]^{-2}du^2.
\end{split}
\end{equation}
 In  (\ref{5dAdSu}) one can easily recognize the 5d AdS metric supported by  the constant dilaton 
\bea\label{phi2b0}
\phi_{u\to u_0}\sim \frac{9k}{(16-9k^2)}\log \frac{3k}{4}+\frac{9k}{2(16-9k^2)}\log \Big|\frac{C_1}{C_2}\Big|,
\eea
which coincides with the minimum of the potential. \\
Introducing a new  conformal "radial" coordinate   as $z =  4(u - u_{0})^{1/4}$ one represents the solution in the Poincare patch 
\bea\label{5dAdS}
ds^{2} \sim \frac{1}{z^{2}}(- dt^{2} + dx^{2}_{1} +dx^{2}_{2}+dx^{2}_{3} + dz^{2}),
 \eea
where  $z\to 0$ as $u\to u_{0}$. The energy scale $e^{\mathcal{A}}$ takes big values with constant dilaton,  see fig.~\ref{fig:scalefactor} {\bf c)}, so this boundary is related to the UV fixed point. 

The asymptotics of the metric and the dilaton with $u \to +\infty$ coincide with those relations for the "right"\ solution  and correspond to the IR-region.

To summarize, we have three holographic RG flows at zero temperature, namely:
\begin{enumerate}
\item the middle solution defined for $u \in (u_{02}, u_{01})$ interpolates between two different hyperscaling violating boundaries that can be associated with the weakly coupled UV fixed point and the IR fixed point at strong coupling;
\item the right solution defined for $u \in (u_{01}, +\infty)$ represents an RG flow with UV and IR fixed points corresponding to a free theory. It also has two different hyperscaling violating boundaries;
\item the special case defined for $u \in (u_{02}, u_{01})$  with $u_{02}=u_{01}=u_{0}$, when the  solution interpolates between AdS-boundary in the UV limit and  hyperscaling violating boundary in the IR limit.
\end{enumerate}
We would like to note that the boundary near $u_{01}$ corresponds to the UV region both holographic RG flows defined on $(u_{01},u_{02})$ and $(u_{0},+\infty)$.

\setcounter{equation}{0}
\section{Wilson loops in holographic RG flow at zero T}\label{Sec:WLT0}
The expectation value of the rectangular temporal Wilson loop of size $T\times \ell$ is related with the potential between a quark and an antiquark on a distance $\ell$ as
\bea
\Braket{W} \sim e^{-V_{q\bar{q}}(\ell) T}.
\eea
At the same time the holographic Wilson loop can be defined through the Nambu-Goto action $\mathcal{S}_{NG}$ of a string in a considered background \cite{Malda, ReyYee}
\bea
\Braket{W(\mathcal{C})}\sim e^{-\mathcal{S}_{NG}}.
\eea
So one can estimate the potential of the quark-antiquark interaction as 
\bea
V_{q\bar{q}} =  \frac{1}{T}\mathcal{S}_{NG}.
\eea

The Nambu-Goto action is defined as
\bea\label{NGdef1}
\mathcal{S}_{NG} = -\frac{1}{2\pi \alpha'}\int d^{2}\sigma\sqrt{-\det{h}},
\eea
where we use the string frame for the induced metric 
\bea\label{NGdef2}
h_{\alpha\beta} = e^{\frac{4}{3}\phi}G_{\mu\nu}\partial_{\alpha}X^{\mu}\partial_{\beta}X^{\nu},
\eea
with the world-sheet coordinated $\sigma^{\alpha}$, $\alpha=0,1$, and the embedding functions $X^{\mu} = X^{\mu}(\sigma^{\alpha})$.
 
In this section we discuss  the Nambu-Goto action for the general solution (\ref{MABCnv}) with $\alpha^{1}=0$ in the domain wall coordinates (\ref{domainwall}). 
This corresponds to consideration of Wilson loops in the holographic RG flows. For the string configuration we choose the following gauge \cite{Malda}
\bea
\sigma^{0} = t, \quad \sigma^{1} = x_{1},\quad  w= w(x_{1}),
\eea
supposing that $w(0) = w(\ell) =0$, $w\left(\frac{\ell}{2}\right) = w_{*}$.

Then  the induced metric has the following form
\begin{equation}
ds^{2}= e^{\frac{4}{3}\phi}\left(-e^{2\mathcal{A}}dt^{2} +  (e^{2\mathcal{A}}+ w'^2)dx^{2}_{1}\right).
\end{equation}

The Nambu-Goto action in the string frame is related to introducing the dilaton multiplier in the following way
\begin{equation}\label{Ss}
\begin{split}
\mathcal{S}_{NG}&=-\frac{1}{2\pi\alpha'}\int dx_1dt e^{\frac{4\phi}{3}}\sqrt{e^{2\mathcal{A}}( e^{2\mathcal{A}}+ w'^2)}.
\end{split}
\end{equation}
The equation for the turning point following from (\ref{Ss}) reads
\bea\label{xprime}
x'_{1} =\frac{c}{e^{\mathcal{A}}\sqrt{e^{4\mathcal{A}+ \frac{8}{3}\phi}-c^{2}}},
\eea
where $x'_{1}= \frac{dx_{1}}{dw}$ and $c$ is a constant, i.e. $c= e^{2\mathcal{A}(w_{*})}$.  
Now it's convenient to come to the original $u$-coordinate  using 
\begin{equation}\label{chcoord}
\begin{split}
x'_{1}&=\frac{dx_{1}}{dw}=\frac{dx_{1}}{du}\frac{du}{dw}=e^{-4\mathcal{A}}\dot{x}_{1},
\end{split}
\end{equation}
where we take into account  (\ref{dwtr})-(\ref{scaleE-dw}) and denote $\dot{x}_{1}=\frac{dx_{1}}{d u}$.

From (\ref{xprime})-(\ref{chcoord}) one can find  the distance between quarks as
\begin{equation}
\begin{split}
\frac{\ell}{2}&=\int du\frac{ce^{3\mathcal{A}}}{\sqrt{e^{4\mathcal{A} +\frac{8}{3}\phi}-c^2}}
\end{split}
\end{equation}
and for the Nambu-Goto action we have the following relation
\begin{equation}\label{NG-sf}
\begin{split}
\frac{\mathcal{S}_{NG}}{2}&=\frac{T}{2\pi\alpha'}\int {du}\frac{e^{7\mathcal{A}+\frac{8\phi}{3}}}{\sqrt{e^{4\mathcal{A}+\frac{8\phi}{3}}-c^2}}.
\end{split}
\end{equation}
Let us define the  so-called effective potential with $u'=0$ \cite{ARS,Arefeva} as
\bea\label{Veff}
V_{eff}  = e^{2\mathcal{A}+\frac{4}{3}\phi} =F_1^{\frac{4(2-3k)}{9k^2-16}}F_2^{\frac{3k(3k-8)}{2(16-9k^2)}}.
\eea
\begin{figure}[t!]
\centering
 \includegraphics[width=6cm]{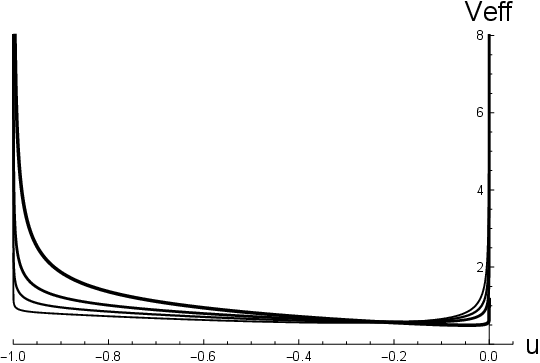}{\bf a} \includegraphics[width=1.5cm]{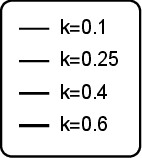}$\,$
 \includegraphics[width=6 cm]{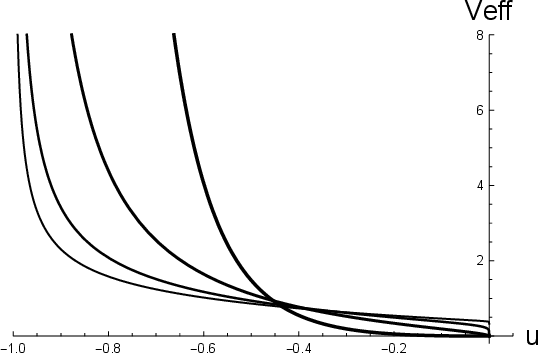}{\bf b} \includegraphics[width=1.5cm]{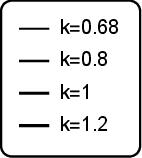}\\
 $\,$\\
  \includegraphics[width=4.8 cm]{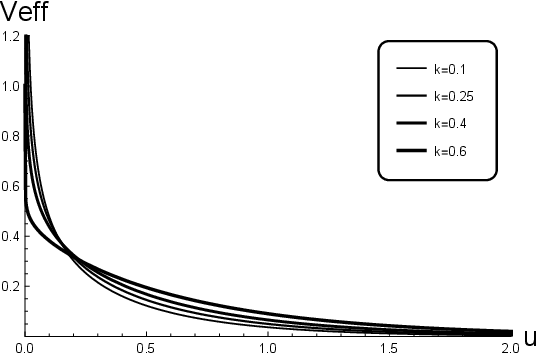}{\bf c}$\,\,$
   \includegraphics[width=4.8 cm]{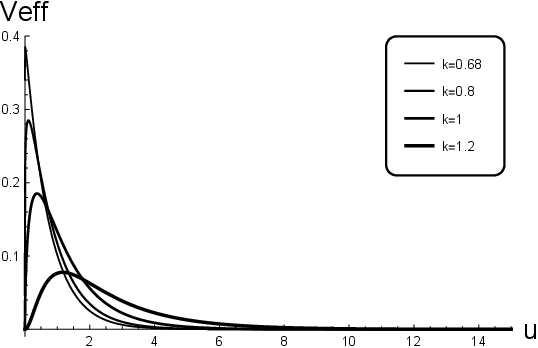}{\bf d}$\,\,$
     \includegraphics[width=4.8 cm]{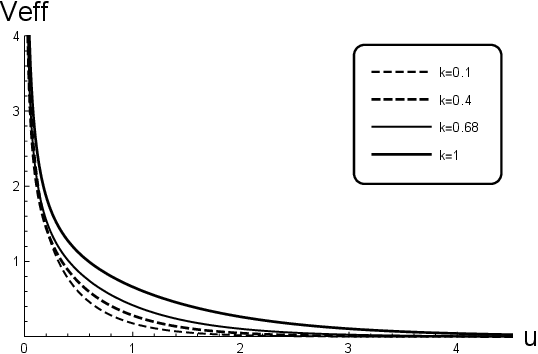}{\bf e}$\,$
   \caption{$V_{eff}$ as a function of $u$ for the holographic RG flows: {\bf a),b)} defined on $(u_{02};u_{01})$ with $u_{02}=-1$, $u_{01} =0$  for small and big $k$, correspondingly; {\bf c),d)} defined on $(u_{01};+\infty)$ with $u_{01} =0$ for small and big $k$, correspondingly; {\bf e)} with coinciding points $u_{01}=u_{02}= 0$. For all $E_{1}= E_{2} = -1$, $C_{1} =- C_{2} = -1$. 
   }
 \label{fig:Veff}
\end{figure}
In terms of $V_{eff}$ the quark-antiquark distance  can be presented as
\bea\label{ellVeff}
\frac{\ell}{2}  =  \int du\, e^{-2\phi}\frac{V_{eff}(u)\sqrt{V_{eff}(u)}}{\sqrt{\frac{V^{2}_{eff}(u)}{V^{2}_{eff}(u_{*})}-1}}
\eea
 and the string action is given by
\bea\label{SVeff}
\frac{\mathcal{S}_{NG}}{2}& =& \frac{T}{2\pi\alpha'}\int {du}\frac{e^{-2\phi}V^{3}_{eff}(u)\sqrt{V_{eff}(u)}}{\sqrt{V^{2}_{eff}(u)- V^{2}_{eff}(u_{*})}}.\eea

One can see from (\ref{ellVeff})-(\ref{SVeff}) that we have to require the function $V_{eff}$ to be decreasing on the region $(u_{01},u_{*})$.
 It can be also seen from (\ref{ellVeff})-(\ref{SVeff}) that to observe a confinement in the IR region, that implies $\ell\to +\infty$ with $\mathcal{S}_{NG}\to +\infty$, $V_{eff}$ needs have a local minimum. 
Let's suppose that
\bea
V'_{eff}(u)|_{u = u_{min}} = 0, \quad V^{''}_{eff}(u_{min})>0.
\eea
Then, expanding in the Taylor series at the point $u_{min}$, with $u_{min} =u_{*}$ one has
\bea\label{Veffexp}
\frac{V^{2}_{eff}(u)}{V^{2}_{eff}(u_{min})} =1  + \frac{V''(u_{min})}{V_{eff}(u_{min})}(u - u_{min})^{2} + \mathcal{O}(u- u_{min})^{2}. 
\eea
We note that the constraint $V'_{eff} = 0$ leads to the equation 
\bea\label{Veffpreq}
2(2-3k)\frac{\cosh(\mu_{1}(u-u_{01}))}{\sinh(\mu_{1}(u-u_{01}))} - (3k -8)\frac{\cosh(\mu_{2}(u-u_{02}))}{\sinh(\mu_{2}(u-u_{02}))} = 0,
\eea
which we need to study separately for each holographic RG flow.

Remembering that the holographic RG flow is splitted into 3 branches one has to analyze the behaviour of (\ref{Veff})-(\ref{Veffpreq}) for  each case of the holographic RG flows.
\begin{enumerate}
\item For holographic RG flow defined on $(u_{02};u_{01})$ ({\bf the middle solution} ) the worldsheet boundary is attached at $u_{01}$. The distance between quarks reads 
\bea\label{midWL1}
\frac{\ell}{2} = \int^{u_{01}}_{u_{*}} du\, \frac{c\, e^{-2\phi}V^{3/2}_{eff}}{\sqrt{V^{2}_{eff}-V^{2}_{eff}(u_{*})}}.
\eea
The string action is divergent at the UV boundary, the renormalized Nambu-Goto action takes the form\footnote{here and what follows we use the approach discussed in \cite{Malda} and generalized in \cite{Gia}}
\bea\label{midWL2}
\mathcal{S}^{\,\,\,ren}_{NG} = \frac{T}{\pi \alpha'}\int^{u_{01}}_{u_{*}}du\, \left(\frac{e^{-2\phi}V^{7/2}_{eff}}{\sqrt{V^{2}_{eff}- V^{2}_{eff}(u_{*})}}- e^{-2\phi}V^{5/2}_{eff} \right) + \frac{T}{\pi \alpha'}\int^{u_{01}-u_{(\Lambda)}}_{u_{*}}du\, e^{-2\phi}V^{5/2}_{eff},\nonumber\\
\eea
where $u_{(\Lambda)}<0$ and the singular term at the boundary $u_{01}$ reads
\bea\label{Srenorm}
\lim_{u\to u_{01}-\epsilon} e^{-2\phi}V^{5/2}_{eff} \sim (\sqrt{\left|\frac{C_{2}}{2E_{2}}\right|}\sinh(\mu_{2}(u_{01}-u_{02})))^{\frac{3k(15k-16)}{4(16-9k^{2})}}(\sqrt{\left|\frac{C_{1}}{2E_{1}}\right|}(\mu_{1}(u-u_{01}))^{\frac{4(5-3k)}{9k^{2} - 16}}.\nonumber\\
\eea
\begin{figure}[h!]
\centering
\includegraphics[width=5cm]{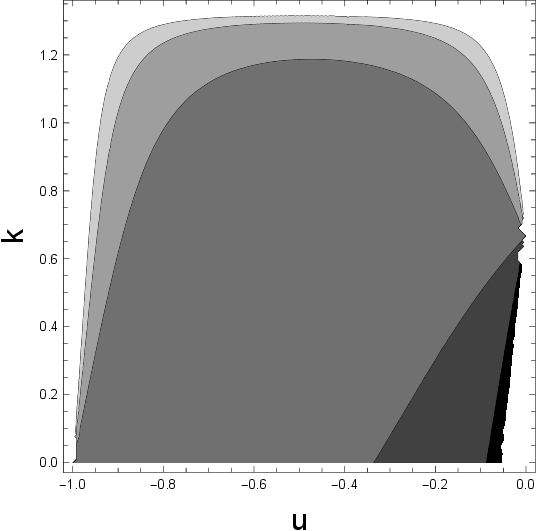}{\bf a}$\,$
   \includegraphics[width=5 cm]{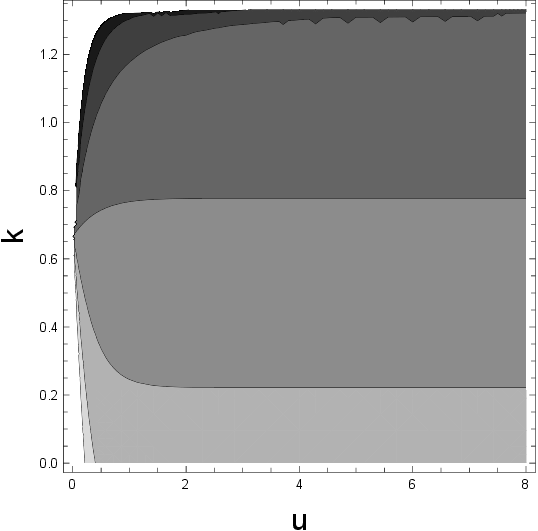}{\bf b}$\,$
\includegraphics[width=5cm]{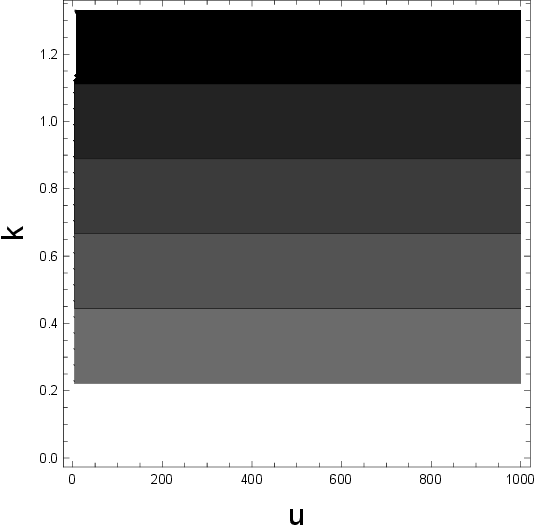}{\bf c}
    \caption{The solutions to the equation (\ref{Veffpreq}) defined on different regions: {\bf a)} $(u_{02},u_{01})$ with $u_{02}=-1$, $u_{01}=0$; {\bf b)} $(u_{01}, +\infty)$, $u_{01}=0$; {\bf c)} $u_{01}=u_{02}=0$. 
   }
 \label{fig:ContPT0}
\end{figure}
By virtue of (\ref{Srenorm}) one can represent (\ref{midWL2}) in the following form
\bea\label{Srenorm2}
&&\mathcal{S}^{ren}_{NG} =\frac{T}{\pi\alpha'} \int^{u_{01}}_{u_{*}}du\, \Bigl(\frac{e^{-2\phi}V^{7/2}_{eff}}{\sqrt{V^{2}_{eff}- V^{2}_{eff}(u_{*})}}- e^{-2\phi}V^{5/2}_{eff} \Bigr) \nonumber\\
 &-&\frac{T}{\pi \alpha'}\frac{2\sqrt{3}}{(2-3k)^{2}\sqrt{|C_{1}|}}(\sqrt{\frac{|C_{2}|}{2|E_{2}|}}\sinh(\mu_{2}(u_{01}-u_{02})))^{\frac{3k(15k-16)}{4(16-9k^2)}}(\sqrt{\frac{|C_{1}|}{2|E_{1}|}}(\mu_{1}|u_{*}-u_{01}|)^{\frac{(3k-2)^{2}}{9k^{2}-16}}.\nonumber\\
\eea

We note that in the UV limit $V_{eff}|_{u\to u_{01}-\epsilon}$ tends $+\infty$ for $k<2/3$ and goes to $0$ for $k>2/3$, while in the IR limit ($u\to u_{02}+ \epsilon$) it always tends to $+\infty$ for all $k$. From figure~\ref{fig:Veff}~{\bf a)} we see that the function $V_{eff}$ has one minimum, for small values $k$ ($k<\frac{2}{3}$). At the same time from figure~\ref{fig:Veff}~{\bf b)} one can observe that for the same RG flow with $k>\frac{2}{3}$  $V_{eff}$ doesn't have any extremal point, so $V'_{eff}\neq 0$. 
Indeed from figure \ref{fig:ContPT0}~{\bf a)} one can see that equation (\ref{Veffpreq}), corresponding to the RG flow on the region $(u_{02};u_{01})$, has a solution for $k<2/3$, while for  $k>\frac{2}{3}$ it does not have solutions. Hereby,
\begin{itemize}
\item for $k<2/3$  taking  into account $V'_{eff}(u_{min})=0$ and plugging (\ref{Veffexp}) in  (\ref{ellVeff})-(\ref{SVeff}) the quark-antiquark distance reads
\bea\label{ellf1}
\frac{\ell}{2}= \int^{u_{01}}_{u_{min}}\frac{e^{-2\phi}V^{2}_{eff}(u_{min})du}{\sqrt{V''_{eff}(u_{min})}(u_{min}-u)}\sim \frac{e^{-2\phi(u_{min})}V^{2}_{eff}(u_{min})}{\sqrt{V''_{eff}(u_{min})}}\log(u_{min}-u),
\eea
and the Nambu-Goto action is given by
\bea\label{NGAf1}
\mathcal{S}_{NG} = \int^{u_{(\Lambda)}}_{u_{min}}\frac{e^{-2\phi}V^{3}_{eff}(u_{min})du}{\sqrt{V''_{eff}(u_{min})}(u_{min}-u)}\sim \frac{e^{-2\phi(u_{min})}V^{3}_{eff}(u_{min})}{\sqrt{V''_{eff}(u_{min})}}\log(u_{min}-u),\nonumber\\
\eea
so $\ell \to +\infty$ with $\mathcal{S}_{NG}\to +\infty$ as $u\to u_{min}$. From (\ref{ellf1})-(\ref{NGAf1}) we have
\bea
\mathcal{S}_{NG} \sim \sigma \ell,
\eea
with $\sigma= V_{eff}(u_{min})$. So at large distances one has a linear growth of the heavy quark potential;
\item if $k>2/3$ we have $V'_{eff}\neq 0$ according to figure~\ref{fig:Veff}~{\bf b)} and we can not reach the regime of large $\ell$. 
\end{itemize}
 Therefore, one has a confinement phase for $k<\frac{2}{3}$ and deconfinement phase for $k>\frac{2}{3}$.
We note that in \cite{AGP} it was shown that the holographic RG flow defined on the region $(u_{02};u_{01})$ reproduces the QCD behaviour of the running coupling.  

\item For the holographic RG flow defined on $(u_{01};+\infty)$  ({\bf the right solution}) the worldsheet boundary is  attached at $u_{01}$ again. The distance between quarks reads 
\bea
\frac{\ell}{2} = \int^{u_{*}}_{u_{01}} du\, \frac{c\, e^{-2\phi}V^{3/2}_{eff}}{\sqrt{V^{2}_{eff}-V^{2}_{eff}(u_{*})}}.
\eea
The Nambu-Goto action is divergent near $u_{01}$ boundary, after the renormalization one has
\bea
&&\mathcal{S}^{\,\,ren}_{NG} =\frac{T}{\pi \alpha'} \int^{u_{*}}_{u_{01}}du\, \left(\frac{e^{-2\phi}V^{7/2}_{eff}}{\sqrt{V^{2}_{eff}- V^{2}_{eff}(u_{*})}}- e^{-2\phi}V^{5/2}_{eff} \right)\nonumber\\ 
 &&+\frac{T}{\pi \alpha'}\frac{2\sqrt{3}}{(2-3k)^{2}\sqrt{|C_{1}|}}(\sqrt{\frac{C_{2}}{2E_{2}}}\sinh(\mu_{2}(u_{01}-u_{02})))^{\frac{3k(15k-16)}{4(16-9k^2)}}(\sqrt{\frac{|C_{1}|}{2|E_{1}|}}(\mu_{1}|u_{*}-u_{01}|)^{\frac{(3k-2)^{2}}{9k^{2}-16}}.\nonumber\\
\eea
 $V_{eff}|_{u\to u_{01}+ \epsilon}$ tends to  $+\infty$ for $k<2/3$ and to $0$ for $k>2/3$ near the $UV$ fixed point.  In the IR limit $V_{eff}$ goes to $0$ for all k.  In figure~\ref{fig:Veff}~{\bf c)} we observe that  $V_{eff}$ doesn't have any extremal points with $k<\frac{2}{3}$ , i.e. $V'_{eff}\neq0$.  For $k>\frac{2}{3}$ figure~\ref{fig:Veff}~{\bf d)} shows that $V_{eff}$ has a maximum, so $V'_{eff}=0$ and $V''_{eff} <0$.   In figure ~\ref{fig:ContPT0}~{\bf b)} we present the solution to eq.(\ref{Veffpreq}) on the region $(u_{01}, +\infty)$.  One can see that there is no solutions  for $k<2/3$, while for $k> 2/3$  there is a solution,  corresponding to maximum of  $V_{eff}$. So, one have the background is non-confining  for all values of $0<k<4/3$. This is in agreement with \cite{AGP},where we have seen  that the RG flow defined on $(u_{01},+\infty)$ corresponds to the UV and IR free theory.

\item The special case ({\bf holographic RG flow with AdS boundary}), $u_{01}=u_{02}=u_{0}$.
 The worldsheet boundary is attached at $u_{0}$. The distance between quarks reads 
\bea
\frac{\ell}{2} = \int^{u_{*}}_{u_{0}} du\, \frac{c\, e^{-2\phi}V^{3/2}_{eff}}{\sqrt{V^{2}_{eff}-V^{2}_{eff}(u_{*})}}.
\eea
The string action is divergent at the UV boundary, the renormalized Nambu-Goto action takes the form
\bea
\mathcal{S}^{ren}_{NG} = \frac{T}{\pi \alpha'}\int^{u_{*}}_{u_{0}}du\, \left(\frac{e^{-2\phi}V^{7/2}_{eff}}{\sqrt{V^{2}_{eff}- V^{2}_{eff}(u_{*})}}- e^{-2\phi}V^{5/2}_{eff} \right) + \frac{T}{\pi \alpha'}\int^{u_{*}}_{u_{0}-u_{(\Lambda)}}du\, e^{-2\phi}V^{5/2}_{eff}.\nonumber\\
\eea
Since we have
\bea
\lim_{u\to u_{0}+\epsilon} e^{-2\phi}V^{5/2}_{eff}\sim |C_{1}|^{\frac{2(5-3k)}{9k^{2}-16}} \left(\frac{4}{3k}\sqrt{C_{2}}\right)^{\frac{3k(15k-16)}{4(16-9k^{2})}}(2E_{1})^{\frac{5}{8}}(\mu_{1}(u- u_{0}))^{-\frac{5}{4}},
\eea
so one can represent the remormalized Nambu-Goto action in the following form
\bea
\mathcal{S}^{\,\,ren}_{NG}&=& \frac{T}{\pi \alpha'}\int^{u_{*}}_{u_{0}}du\, \left(\frac{e^{-2\phi}V^{7/2}_{eff}}{\sqrt{V^{2}_{eff}- V^{2}_{eff}(u_{*})}}- e^{-2\phi}V^{5/2}_{eff} \right) \nonumber\\
&-&\frac{T}{\pi \alpha'}|C_{1}|^{\frac{2(5-3k)}{9k^{2}-16}} \left(\frac{4}{3k}\sqrt{C_{2}}\right)^{\frac{3k(15k-16)}{4(16-9k^{2})}}(2E_{1})^{\frac{5}{8}}\frac{4\mu^{-\frac{5}{4}}_{1}}{(u_{*} -u_{0})^{1/4}}.
\eea
We note that $V_{eff}$ tends to $+\infty$ as $u \to u_{0} +\epsilon$ and to $0$ with $u\to+\infty$ for all $k$.

Figure~\ref{fig:Veff}  ~{\bf e)}  shows that  $V_{eff}$ corresponding to this RG flow  doesn't have any extremal points defined on $(u_{0};+\infty)$  for all values of $0<k<4/3$. This is confirmed by solutions to eq.~(\ref{Veffpreq}) in figure \ref{fig:ContPT0}~{\bf c)}. So the holographic RG flow with an AdS UV fixed point is also non-confining. It is consistent with an observation from \cite{AGP}, where it was shown that the running coupling corresponding to this RG flow tends to some finite value in the UV limit and to $0$ in the IR region.

\end{enumerate}

\setcounter{equation}{0}
\section{Wilson loops in the holographic RG flow at finite temperature.}\label{Sec:WLTf}
Now we turn to the discussion of the Wilson loops in the non-zero temperature background found in \cite{AGP}. 
It was shown that   one can derive the black brane metric from the generic non-Poincar\'e invariant solutions with non-zero $\alpha^{1}$ (\ref{MABCnv})-(\ref{E1E2c3}) defined on the region $(u_{01};+\infty)$\footnote{without loss of generality we put $u_{01}=0$ to obey $f=1$ near $u_{01}$}. It is reached taking $\mu_{1} = \mu_{2} = -\frac{4}{3}\alpha^{1}$, that is possible for the non-Poincar\'e invariant solutions by virtue of the constraint (\ref{E1E2c3}).  The black hole metric reads
   \bea\label{f}
ds^{2}
& = & \mathcal{C}\,\mathcal{X}(u)\Big( - e^{-2\mu u}dt^{2} + d\vec{x}^{2} \Big)+ 
 \mathcal{C}^4\,\mathcal{X}(u)^4
e^{-2\mu u}du^{2},\label{1.3mm}
\eea
where $\mathcal{X}$ and $\mathcal{C}$ are given by
\bea\label{Xbbxbb}
\mathcal{X} &=&(1-e^{-2\mu u})^{-\frac{8}{16-9k^{2} }}(1-e^{-2\mu (u-u_{02})})^{\frac{9k^{2}}{2(16- 9k^{2})}},\\ \label{cbbxbb}
\mathcal{C}&\equiv& 2^{\frac{16}{(16-9k^{2})}} (3 \mu)^{\frac{1}{2}}\left|C_{1}\right|^{\frac{8}{2(9k^{2} - 16)}}
\Big(\frac{C_{2}}{k}e^{-2\mu u_{02}}\Big)^{\frac{9k^{2}}{4(16- 9k^{2})}} (16-9k^{2})^{-\frac{1}{4}}
\eea
and the corresponding dilaton is
\bea\label{dilaton-bb}
\phi  = \frac{9k}{9k^{2}-16}\log{\left[\frac{4}{3k}\sqrt{\left|\frac{C_{2}}{C_{1}}\right|}\frac{\sinh(\mu(u-u_{02}))}{\sinh(\mu u)}\right]},
\eea
which is convenient to represent in the following form
\bea
e^{\phi} = \left(\frac{4e^{-\mu u_{02}}}{3k}\sqrt{\frac{C_{2}}{|C_{1}|}}\right)^{\frac{9k}{9k^{2} -16}}\left(1-e^{-2\mu(u- u_{02})} \right)^{\frac{9k}{9k^{2} -16}}\left(1-e^{-2\mu u} \right)^{-\frac{9k}{9k^{2} -16}}.
\eea
The horizon of this black brane  is located at $u=+\infty$ and we reach the boundary at $u= 0$  with $f=1$, i.e. the black brane is defined for $u\in (0; + \infty)$ with $u_{01} = 0$. \\
The Hawking temperature is given by
\bea
T=\frac{1}{2\pi}\frac{\mu}{\mathcal{C}^{3/2}}.
\eea
It is worth to be noting that the finite temperature generalization of the holographic  RG flow with an AdS boundary leads to the 5d AdS black hole \cite{AGP}. 

\begin{figure}[t!]
\centering
 \includegraphics[width=5cm]{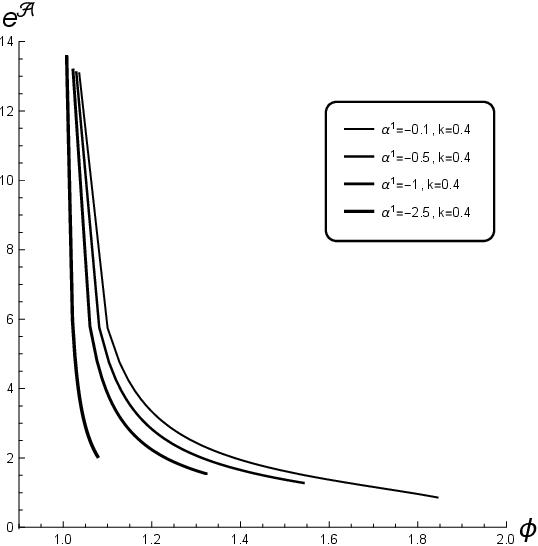}$\,\,${\bf a}$\,\,$
 \includegraphics[width=5cm]{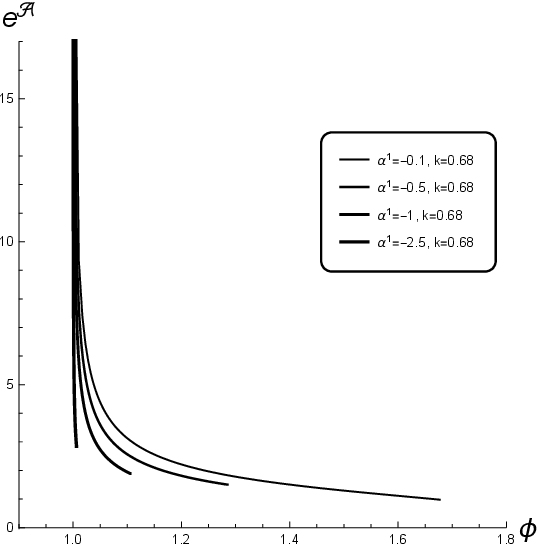}$\,\,${\bf b}$\,\,$
 \caption{The behaviour of $e^{\mathcal{A}}$ as a function on $\phi$ for black holes defined on $(u_{0},+\infty)$: {\bf a)} $k<2/3$ is fixed ($k=0.4$) and $\alpha^{1}=-\frac{3}{4}\mu$ is varying;  {\bf b)} $k>2/3$ is fixed ($k=0.68$) and $\alpha^{1}$ is varying. For both plots we take $C_{1}= -C_{2}=1$.}
  \label{fig:scalefactorT}
\end{figure}

 In \cite{AGP} it was shown that the solution (\ref{f})-(\ref{dilaton-bb}) respects the Gubser bound \cite{Gubser}.
 We see in figure~\ref{fig:scalefactorT} that $e^{\mathcal{A}}$ is monotonic on $\phi$ and, hereby, it can be considered as a holographic RG flow at finite temperature. From fig.~\ref{fig:scalefactorT} one can read  that $e^{\mathcal{A}}$ tends to $+\infty$ at some some small value of the dilaton corresponding the value of the dilaton on the boundary $u =0$. So $u =0$ is associated with the UV fixed point. In fig.~\ref{fig:scalefactorT} we see that we can not reach the deep infrared region.

Now let us turn to the discussion of holographic Wilson loops in the background (\ref{f})-(\ref{dilaton-bb}).
As for the zero-temperature case we consider the rectangular time-like Wilson loop of size $T\times \ell$. The world-sheet boundary is attached at $u=0$.
We choose the following gauge for the string
\be\label{gauge-bb}
\sigma^{0} = t,\quad \sigma^{1} =x_{1}, \quad u = u(x_{1}).
\ee
The induced metric calculated with respect to (\ref{NGdef2}) reads
 \begin{equation}\label{ind-bb}
 \begin{split}
ds^{2} =  e^{\frac{4}{3}\phi}\left(-\mathcal{C}\mathcal{X}e^{-2\mu u}dt^{2} + \mathcal{C}\mathcal{X}(1+\mathcal{C}^3\mathcal{X}^3e^{-2\mu u}u'^2)dx^{2}_{1}\right),
\end{split}
 \end{equation}
where we denote $u' =\frac{du}{dx_{1}}$.

With (\ref{gauge-bb}-\ref{ind-bb}) the Nambu-Goto action (\ref{NGdef1}) can be represented as
\begin{equation}\label{NGATS}
\mathcal{S}_{NG}=-\frac{T}{2\pi \alpha'}\int{dx_1}e^{\frac{4\phi}{3}}\mathcal{C}\mathcal{X}e^{-\mu u}\sqrt{1+\mathcal{C}^3\mathcal{X}^3e^{-2\mu u}u'^2}.
\end{equation}
The equation of motion for the turning point that follows from (\ref{NGATS}) reads
\begin{equation}
u'=\pm\frac{\sqrt{ \mathcal{C}^2\mathcal{X}^2 e^{\frac{8\phi}{3}}e^{-2\mu u}-c^2}}{c\mathcal{C}^{3/2}\mathcal{X}^{3/2}e^{-\mu u}}.
\end{equation}
The distance between  two quarks is 
\begin{equation}
\frac{\ell}{2}=\int^{u_{*}}_{0}du\frac{c\mathcal{C}^{3/2}\mathcal{X}^{3/2}e^{-\mu u}}{\sqrt{\mathcal{C}^2\mathcal{X}^2 e^{\frac{8\phi}{3}}e^{-2\mu u}-c^2}}.
\end{equation}
We consider the case when the turning point $u_{*}<u_{h}$.

The effective potential  can be introduced as follows 
\bea
V_{eff}&=&\mathcal{C}\mathcal{X}e^{-\mu u}e^{\frac{4}{3}\phi} \nonumber\\
&=&\mathcal{C} e^{-\mu u}  \left(\frac{4}{3k}\sqrt{\frac{C_{2}}{|C_{1}|}}\right)^{\frac{12k}{9k^{2} -16}}\left(e^{-\mu u_{02}}\right)^{\frac{12k}{9k^{2}-16}}\left(1-e^{-2\mu(u- u_{02})} \right)^{\frac{3k(8-3k)}{2(9k^{2}-16)}}\left(1-e^{-2\mu u} \right)^{\frac{4(2-3k)}{9k^{2}-16}}.\nonumber\\
\eea
The distance between quarks and the Nambu-Goto action  can represented in terms of $V_{eff}$ as
\bea\label{ellT}
\frac{\ell}{2} = \int^{u_{*}}_{0}du\,\frac{e^{-2\phi}e^{\frac{\mu}{2}u}V_{eff}\sqrt{V_{eff}}}{\sqrt{\frac{V^{2}_{eff}(u)}{V^{2}_{eff}(u_{*})}-1}}
\eea
and
\bea\label{SNGT}
\mathcal{S}_{NG} =\frac{T}{\pi\alpha'} \int^{u_{*}}_{0}du\,\frac{e^{-2\phi}e^{\frac{\mu}{2}u}V^{3}_{eff}\sqrt{V_{eff}}}{\sqrt{V^{2}_{eff}(u) -V^{2}_{eff}(u_{*})}},
\eea
correspondingly.  The string action (\ref{SNGT}) is divergent near the UV fixed point $u=0$.
The singular term near $0$ reads
\bea
\lim_{u\to 0} \mathcal{C}^{5/2}\mathcal{X}^{5/2}e^{\frac{4}{3}\phi}f\sim \mathcal{C}^{5/2}\left(\frac{4}{3k}\sqrt{\frac{C_{2}}{|C_{1}|}}e^{-\mu u_{02}}\right)^{\frac{12k}{9k^{2}-16}}\left(1 - e^{2\mu u_{02}}\right)^{\frac{3k(15k -16)}{4(16-9k^{2})}}(2\mu u)^{\frac{4(3k-5)}{16-9k^{2}}},
\eea
so one can represent the renormalized Nambu-Goto action in the following form
\bea
\mathcal{S}^{ren}_{NG} &= &\frac{T}{\pi\alpha'} \int^{u_{*}}_{0}du\,\left(\frac{e^{-2\phi}e^{\frac{\mu}{2}u}V^{3}_{eff}\sqrt{V_{eff}}}{\sqrt{V^{2}_{eff}(u) -V^{2}_{eff}(u_{*})}} - e^{-2\phi}V^{5/2}_{eff}f\right)\nonumber\\
&+&\frac{T}{\pi \alpha'}\frac{\mathcal{C}^{5/2}}{2\mu}\frac{9k^{2}-16}{(2-3k)^{2}}\left(\frac{4}{3k}\sqrt{\frac{C_{2}}{|C_{1}|}}e^{-\mu u_{02}}\right)^{\frac{12k}{9k^{2}-16}}\left(1 - e^{2\mu u_{02}}\right)^{\frac{3k(15k -16)}{4(16-9k^{2})}}(2\mu u_{*})^{\frac{(3k-2)^{2}}{9k^{2}-16}}.\nonumber\\
\eea

As for the case of zero temperature,  one can see from (\ref{ellT})-(\ref{SNGT}) that $V_{eff}$ should be a decreasing function on $0<u<u_{*}$. To  guarantee the presence of the confinement phase  one needs $V'_{eff}=0$ at the point $u_{min}$. This condition gives rise to the following equation
\bea\label{VeffprimeTeq}
 \left(\frac{4(2-3k)}{\sinh(\mu u)} + \frac{3k(8-3k)}{2\sinh(\mu(u-u_{02}))}+ 16-9k^{2}\right)V_{eff} =0,
\eea
for $u\in (0,+\infty)$.
 
We also note that near the UV boundary  $V_{eff}|_{u\to 0+\epsilon} \to +\infty$ with $k< 2/3$ and tends to $0$ with $k>2/3$, at the same $V_{eff}$ goes to $0$ as $u \to +\infty$ (the IR boundary). 
\begin{figure}[t!]
\centering
\includegraphics[width=5.5cm]{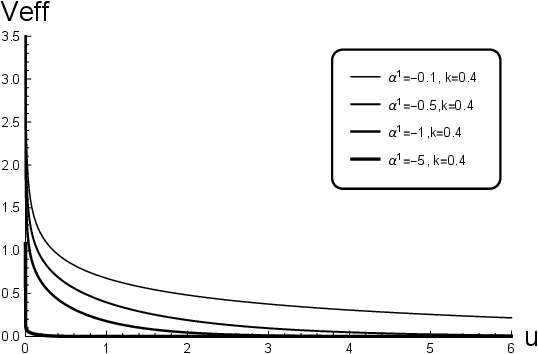}{\bf a}$\,\,\,\,\,$\includegraphics[width=5.5cm]{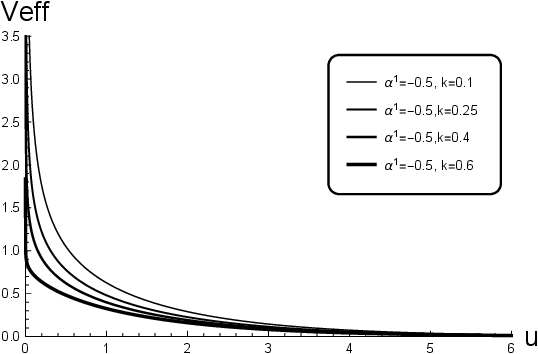}{\bf b}$\,\,\,\,\,$$\,\,\,\,\,$$\,\,\,\,\,$\\ \includegraphics[width=6 cm]{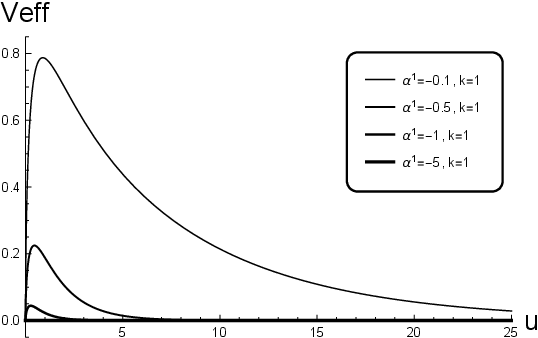}{\bf c}$\,\,\,\,\,$
 \includegraphics[width=6 cm]{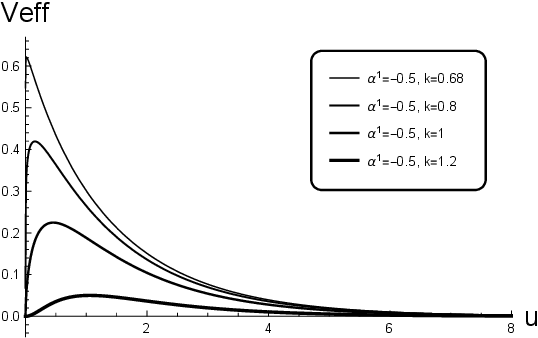}{\bf d}$\,\,\,\,\,$$\,\,\,\,\,$$\,\,\,\,\,$
    \caption{ $V_{eff}$ as a function of $u$ for the holographic RG flows at finite temperature: {\bf a),c)} we fix $k$ varying $\alpha^{1}=-\frac{3}{4}\mu$, {\bf b),d)} we fix $\alpha^{1}=-\frac{3}{4}\mu$ varying $k$.  For the left panel we put $k<2/3$ and for the right panel $k>2/3$. For all we take $C_{1} =- C_{2} = -1$. 
   }
 \label{fig:VeffT}
\end{figure}

\begin{figure}[h!]
\centering
\includegraphics[width=5.5cm]{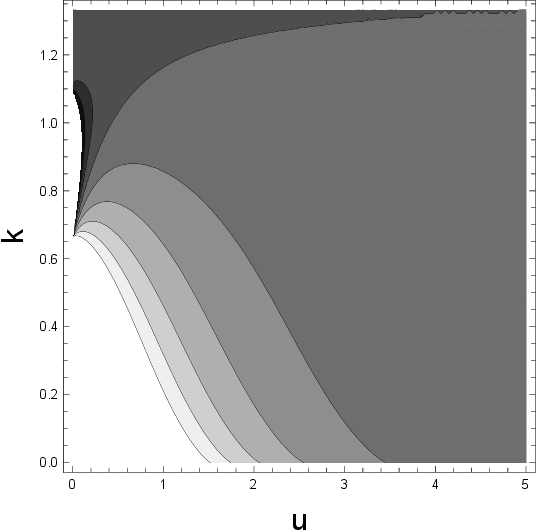}
    \caption{The solutions to the equation (\ref{VeffprimeTeq}) defined on the region $(0; +\infty)$.}
 \label{fig:ContPT}
\end{figure}

In figures~\ref{fig:VeffT} {\bf a),b)} we show the dependence of $V_{eff}$ on $u$ for the thermal case for values of $k<\frac{2}{3}$. In fig.~\ref{fig:VeffT}  {\bf a)} we keep the dilaton coupling as $k=0.4$ and vary the value of the parameter $\alpha^{1} = -\frac{3}{4}\mu$ related to the temperature,  while in fig.~\ref{fig:VeffT} {\bf b)} we change the shape of the dilaton potential at fixed temperature related to the parameter $\alpha^{1}$. For both figures~\ref{fig:VeffT} {\bf a)} and {\bf b)} we see that $V_{eff}$ is monotonically decreasing function and  doesn't have any minimum.
In figures~\ref{fig:VeffT} {\bf c),d)} the behaviour of  $V_{eff}$ for the thermal case with $k>\frac{2}{3}$ is presented. Again, we fix the parameter $k$ changing $\alpha^{1}$ in fig.~{\bf c)} and, vise versa, in fig.~{\bf d)} one  varies $k$ at fixed $\alpha^{1}$. We observe that $V_{eff}$ for $k>2/3$ has a maximum, i.e. $V''_{eff}<0$. In figure~\ref{fig:ContPT} we present   solutions to eq. (\ref{VeffprimeTeq}). We see that there is a solution for $k>\frac{2}{3}$ corresponding to a maximum of $V_{eff}$.  We also see that at high temperatures and  $k\sim1$ the function $V_{eff}$ is almost flat as it can be seen from fig.~\ref{fig:VeffT}~{\bf c)}. So, the behaviour of $V_{eff}$ demonstrates that the thermal RG flow is non-confining.
 It is interesting to note that in \cite{AGP} we observed the Hawking-Page phase transition for $k>\frac{2}{3}$. However, the phase diagram $(\mathcal{F},T)$ cannot be completed since the free energy would be discontinuous at $T_{max}$. 

\section{Conclusion}

In this paper we have analyzed the  behaviour of Wilson loops at long distances in the holographic renormalization group flows constructed in \cite{AGP}. These backgrounds are domain wall and black hole solutions to a 5d dilaton gravity model with an exponential potential. The solutions depend on the parameters $u_{01}$ and $u_{02}$, that yield to different RG flows defined on regions $(u_{01},u_{02})$, $(u_{01},+\infty)$, interpolating between boundaries with hyperscaling violation,  and a special case defined on $(u_{0},+\infty)$ ($u_{01}=u_{02}$) with an AdS boundary.
We have analyzed holographic rectangular time-like Wilson loops at zero and finite temperatures. 

In the case of $T=0$ we have shown that the quark-antiquark potential grows linearly at large $\ell$ for the holographic RG flow defined on domain $(u_{01},u_{02})$, so there is a confinement phase  for certain values of the dilaton coupling constant. This is consistent with the observation from \cite{AGP}, where it was seen that the dependence of the running coupling $\lambda = e^{\phi}$ on the energy scale $e^{\mathcal{A}}$ for this RG flow mimics the QCD behaviour. The long distance Wilson loops calculated in two others holographic RG flows at zero temperature have shown that these backgrounds are non-confining. It confirms the observations from \cite{AGP} where it was pointed the corresponding running couplings of these theories are both UV and IR free.

In the thermal case we have shown that the corresponding RG flow background is in deconfined phase.
However, it should be noted that  the dependence of the energy scale on the dilaton (see fig.~\ref{fig:scalefactorT}) has been indicated that we can not achieve the deep IR.

It would  be interesting to perform more careful analysis of the quark-antiquark potential for the confining case.
In particular,  it is relevant to find out if there any possibility for tuning of the background to reproduce the Cornell potential.
Another relevant issue is to compute the Regge spectrum of the holographic mesons.
It is also interesting to apply stability analysis to Wilson loop configurations as it was done in \cite{Stab}.

 
\section*{Acknowledgments}

We would like to thank I. Ya. Aref'eva, G. Policastro, K. Rannu and P. Slepov for helpful discussions and comments. 
AG is supported by RFBR according to the research project  18-02-40069 mega.

\appendix
\setcounter{equation}{0} \renewcommand{\theequation}{A.\arabic{equation}}

\section{Expansion near $u_{02}$ and $u_{01}$}\label{appendix}
Here  we discuss the holographic RG flow   (\ref{MABCnv})-(\ref{FD.4nv}) with $\alpha^{1}=0$ defined on the region $(u_{02},u_{01})$.\\

$\bullet$ Let us perform expansion of $\mathcal{A}$ end $\phi$ near $u_{02}$.\\
 The scale factor $\mathcal{A}$ (\ref{scaleE-dw}) near $u_{02}$  is
\bea\label{Aexpu02new}
\mathcal{A}&\sim& \frac{4}{(9k^{2} -16)}\log a_{0}- \frac{9k^{2}}{4(9 k^2-16)}\log(a_{1}(u-u_{02}))  +\mathcal{O}((u-u_{02})^2),
\eea
with
\bea
a_{0} = \sqrt{| \frac{C_1}{2E_1}| } \sinh ( \sqrt{| \frac{3E_{1}}{2} (k^2-\frac{16}{9})| }(u_{01}-u_{02})), \quad a_{1}=\frac{2}{3}\sqrt{|C_2 \frac{9k^2-16}{3k^2}|}.
\eea

For the expansion of the dilaton near the point $u_{02}$ we have
\bea\label{dilExp}
\phi &\sim& -\frac{9k}{9k^2-16}\log a_{0} + \frac{9k}{9 k^2-16} \log (a_{1}(u-u_{02}))+\mathcal{O}((u-u_{02})^2).
\eea

Combing the expression for $\mathcal{A}$ and $\phi$ one has
\bea\label{exprAphi1}
e^{3\mathcal{A} }&\sim&a^{\frac{12}{9k^{2} -16}}_{0}(a_{1}(u-u_{02}))^{-\frac{27k^{2}}{4(9k^{2}-16)}}, \\
e^{4\mathcal{A}+\frac{8}{3}\phi}&\sim& a^{\frac{8(2-3k)}{9k^2-16}}_{0}\left(a_{1}(u-u_{02})\right)^{\frac{3k(8-3k)}{9k^2-16}}, \\ \label{exprAphi3}
e^{7\mathcal{A} + \frac{8}{3}\phi}&\sim&  a^{\frac{4(7 -6k)}{9k^{2}-16}}_{0}(a_{1}(u-u_{02}))^{\frac{3k(32-21k)}{4(9k^{2}-16)}}.
\eea
Plugging (\ref{exprAphi1})-(\ref{exprAphi3}) in (\ref{midWL1})-(\ref{Srenorm2}) and integrating one obtains
\bea
\frac{\mathcal{S}_{NG}}{2}& =& \frac{ a^{\frac{4(7 -6k)}{9k^{2}-16}}_{0}(9 k^2-16)(a_{1}(u_{*}-u_{02}))^{\frac{3 (32-21 k) k}{36 k^2-64}+1}
   \sqrt{ a^{\frac{8(2-3k)}{9k^2-16}}_{0}(a_{1}(u_{*}-u_{02}))^{\frac{3 (8-3 k) k}{9 k^2-16}}-c} \,}{c^{2}(3 k-8) (9 k-8)}\nonumber\\
 &\times&  {\bf _2F_1}\left(1,\frac{5}{4}-\frac{2}{3k};\frac{7}{4}-\frac{2}{3 k};\frac{1}{c}  a^{\frac{8(2-3k)}{9k^2-16}}_{0}(a_{1}(u_{*}-u_{02}))^{\frac{3 (8-3 k) k}{9 k^2-16}}\right),
\eea
while the quark-antiquark distance reads
\bea
\frac{\ell}{2}&=&\frac{4 (9 k^2-16)}{c(9 k^2-64)} a^{\frac{12}{9k^{2}-16}}_{0} (a_{1}(u_{*}-u_{02}))^{\frac{27 k^2}{64-36 k^2}+1} \sqrt{a^{\frac{8(2-3k)}{9k^2-16}}_{0} (a_{1} (u_{*}-u_{02}))^{\frac{3 (8-3 k) k}{9 k^2-16}}-c} \,\nonumber\\
   &\times& {\bf_2F_1}\left(1,\frac{1}{4}-\frac{2}{3k};\frac{3}{4}-\frac{2}{3 k};\frac{a^{\frac{8(2-3k)}{9k^2-16}}_{0}}{c}  (a_{1} (u_{*}-u_{02}))^{\frac{3 (8-3 k) k}{9 k^2-16}}\right).
\eea
$\bullet$  The expansion near $u_{01}$ of the scale factor and the dilaton read
\bea
\mathcal{A}& \sim&  \frac{4}{(9k^{2} -16)}\log (b_{0}(u-u_{01})) - \frac{9 k^2}{4(9 k^2-16)}\log b_{1} +\mathcal{O}((u-u_{01})^2), \\
\mathcal{\phi} &\sim&-\frac{9k}{9 k^2-16}\log(b_{0}(u-u_{01}))+  \frac{9 k}{9 k^2-16}\log b_{1}+\mathcal{O}((u-u_{01})^2),
\eea
where
\bea
b_{0} =\frac{1}{2}\sqrt{3| C_{1} (k^2-\frac{16}{9})| }, \quad b_{1}=\sqrt{| \frac{C_{2}}{2E_{2}}| } \sinh
   (\frac{2}{3} \sqrt{\frac{2}{3}} u_{02} \sqrt{| E_{2}
   (\frac{16}{k^2}-9)| }).
\eea
Then we have
\bea\label{expaphiu011}
e^{3\mathcal{A} }&\sim&(b_{0}(u-u_{01}))^{\frac{12}{9k^{2} -16}}b_{1}^{-\frac{27k^{2}}{4(9k^{2}-16)}}, \\
e^{4\mathcal{A}+\frac{8}{3}\phi}&\sim& (b_{0}(u-u_{01}))^{\frac{8(2-3k)}{9k^2-16}}b_{1}^{\frac{3k(8-3k)}{9k^2-16}}, \\ \label{expaphiu012}
e^{7\mathcal{A} + \frac{8}{3}\phi}&\sim& (b_{0}(u-u_{01}))^{\frac{4(7 -6k)}{9k^{2}-16}}b_{1}^{\frac{3k(32-21k)}{4(9k^{2}-16)}}.
\eea
Inserting (\ref{expaphiu011})-(\ref{expaphiu012}) in the expression for $\ell$ and integrating one gets (\ref{midWL1})
\bea
\frac{\ell}{2}& =& -\frac{9 k^2-16}{(9 k^2-4)b_{0}}b_{1}^{-\frac{27k^{2}}{4(9k^{2}-16)}}(b_{0}(u_{*}-u_{01}))^{\frac{12}{9 k^2-16}+1}
   \sqrt{ b_{1}^{\frac{3k(8-3k)}{9k^2-16}} (b_{0}(u_{*}-u_{01}))^{-\frac{8 (3 k-2)}{9 k^2-16}}-1} 
  \nonumber\\
  &\times& { \bf _2F_1}\left(1,\frac{(2-3 k)}{8} ;-\frac{3(k-2)}{8} ;\frac{1}{c^{2}} b_{1}^{\frac{3k(8-3k)}{9k^2-16}}
   (b_{0}(u_{*}-u_{01}))^{\frac{8 (2-3 k)}{9 k^2-16}}\right).
\eea
Doing the same for the string action $\mathcal{S}_{NG}$ (\ref{Srenorm2}) we have
\bea
\mathcal{S}_{NG}& =& -\frac{ (9 k^2-16) b_{0}^{\frac{4(7 -6k)}{9k^{2}-16}} b_{1}^{\frac{3k(32-21k)}{4(9k^{2}-16)}}}{3 (k-2) (3 k-2) c}(u_{*}-u_{01})^{\frac{3 (k-2) (3 k-2)}{9
   k^2-16}} \sqrt{\frac{1}{c^{2}} b_{1}^{\frac{3k(8-3k)}{9k^2-16}} (b_{0}(u_{*}-u_{01}))^{-\frac{8 (3 k-2)}{9 k^2-16}}-1} \,
 \nonumber\\
&\times&  {\bf _2F_1}\left(1,\frac{5}{4}-\frac{3 k}{8};\frac{7}{4}-\frac{3 k}{8}; \frac{1}{c^{2}} b_{1}^{\frac{3k(8-3k)}{9k^2-16}}
   (b_{0}(u_{*}-u_{01}))^{\frac{8 (2-3 k)}{9 k^2-16}}\right).
\eea

\end{document}